\documentclass[10pt,twocolumn,letterpaper]{article}

\usepackage{cvpr}
\usepackage{times}
\usepackage{epsfig}
\usepackage{graphicx}
\usepackage{amsmath}
\usepackage{amssymb}
\usepackage{indentfirst}
\usepackage{float}
\usepackage{booktabs}
\usepackage{accents}
\usepackage{relsize}

\usepackage[pagebackref=true,breaklinks=true,letterpaper=true,colorlinks,bookmarks=false]{hyperref}

\def\cvprPaperID{7467} 

\ifcvprfinal\fi
\begin{document}

\title{SAUNet: Shape Attentive U-Net for Interpretable Medical Image Segmentation}

\author{\large Jesse Sun$^{1,4}$, Fatemeh Darbehani$^{1,2,3}$, Mark Zaidi$^{2}$, Bo Wang$^{1,2,3}$\\
\normalsize Peter Munk Cardiac Center$^{1}$, University of Toronto$^{2}$, Vector Institute$^{3}$, University of Waterloo$^{4}$\\
{\tt\small j294sun@uwaterloo.ca, \{fatemeh.darbehani, mark.zaidi\}@mail.utoronto.ca}
\\ \tt\small bowang@vectorinstitute.ai
}

\maketitle

\begin{abstract}
   Medical image segmentation is a difficult but important task for many clinical operations such as cardiac bi-ventricular volume estimation. More recently, there has been a shift to utilizing deep learning and fully convolutional neural networks (CNNs) to perform image segmentation that has yielded state-of-the-art results in many public benchmark datasets. Despite the progress of deep learning in medical image segmentation, standard CNNs are still not fully adopted in clinical settings as they lack robustness and interpretability. Shapes are generally more meaningful features than solely textures of images, which are features regular CNNs learn, causing a lack of robustness. Likewise, previous works surrounding model interpretability have been focused on post hoc gradient-based saliency methods. However, gradient-based saliency methods typically require additional computations post hoc and have been shown to be unreliable for interpretability. Thus, we present a new architecture called Shape Attentive U-Net (SAUNet) which focuses on model interpretability and robustness. The proposed architecture attempts to address these limitations by the use of a secondary shape stream that captures rich shape-dependent information in parallel with the regular texture stream. Furthermore, we suggest multi-resolution saliency maps can be learned using our dual-attention decoder module which allows for multi-level interpretability and mitigates the need for additional computations post hoc. Our method also achieves state-of-the-art results on the two large public cardiac MRI image segmentation datasets of SUN09 and AC17.
\end{abstract}

\section{Introduction}
\indent Cardiovascular magnetic resonance imaging (CMR) is currently used as the gold standard for the non-invasive assessment of various cardiovascular functions \cite{RN225,RN220}. The high spatial resolution and absence of ionizing radiation during CMR confers several advantages over nuclear medicine imaging modalities such as computed tomography (CT), positron emission tomography (PET), and single-photon emission computed tomography (SPECT) \cite{RN217}. As a result, CMR plays a crucial role in the diagnosis of cardiovascular diseases such as cardiomyopathies, myocarditis, and congenital heart disease \cite{RN218}. Diagnosis of such diseases requires specific measurements of morphological and intensity-based features in the image. For example, measurements of the left and right ventricular volumes, along with heart rate, can be used to quantify cardiac output. An increase in cardiac output at resting state is associated with the development of cardiovascular disease caused by atherosclerosis, whereas a low cardiac output can be indicative of heart failure and cardiomyopathy. There are numerous diseases that require bi-ventricular volume estimation in disease diagnosis and prognosis \cite{RN221,RN222,RN219}.

\indent Arguably one of the greatest challenges in bi-ventricular volume estimation is the segmentation of the left and right ventricular endocardium at end-systolic and diastolic timepoints. In a clinical setting, there is a high reliance on manual annotations for ventricular volume estimation. CMR accredited operators will typically annotate short-axis slices by drawing a polygon around the left ventricle (LV) on slices that contain it. The inclusion criteria for which slices to annotate the LV on is ambiguous and can range from slices containing at least 50 to 75\% of the cavity being surrounded in the myocardium. Furthermore, if a slice contains both ventricle and atrial myocardium, the operator may choose to trace through the junction in a straight line. Since detailed contouring is very time consuming, one method to annotate the LV includes drawing a circle around the endocardium and performing binary thresholding to create a mask of all pixels with an intensity comparable to that of blood. In either case, there is a heavy reliance on a human operator to calculate LV volume. Geometrical models have been developed in an attempt to decrease the time spent performing annotations, however both manual annotations and geometric modeling suffer from poor intra-observer and inter-observer variability \cite{RN209}. Recent advances in imaging and computing have led to a drastic rise in the use of machine learning for medical imaging \cite{RN210}. The advent of deep learning allows for much higher levels of abstraction for feature selection and discretization. Convolutional neural networks (CNNs) have been shown to learn abstractions obtained from multidimensional medical images, learning features hard to define by humans. This is one of the reasons why CNNs excel at object recognition and segmentation \cite{RN216}.
\smallskip
\newline
\indent As a result, image segmentation methods in medical imaging have diverted to deep learning solutions. Ronneberger et al. \cite{RN219} proposed the U-Net architecture which have gained a lot of traction in medical imaging. The appeal of the U-Net architecture is the dual contracting and symmetric expanding path connected via skip connections to capture context and localization, respectively \cite{RN226}. Unsurprisingly, the versatility of U-Nets has been demonstrated much promise for cell segmentation and tracking. In the ISBI cell tracking challenge in 2015 and the EM segmentation challenge at ISBI 2012, U-Nets came in first place, outperforming DIVE and IDSIA networks \cite{RN226,RN212}. More recently, U-Net inspired architectures have been employed for segmentation of cardiac structures in CMR images \cite{RN208}. Using the AC17 dataset \cite{RN225}, segmentation of the left and right ventricular cavity (LVC, RVC), and left ventricular myocardium (LVM) were performed at each frame of the cardiac cycle. Dice scores of 0.945 (LVC), 0.908 (RVC), and 0.905 (LVM) were achieved in a training set cross-validation \cite{RN208,RN211}. Furthermore, by permuting the training set with random rotations, mirroring, elastic deformations, and slice misalignments, the resulting model became robust against slice misalignments, cardiac diseases, and ventricular deformations \cite{RN208}.
\newline
\indent One disadvantage shared by many neural networks including U-Net is a lack of interpretability. Because these neural networks interface with many convolutional layers simultaneously, it becomes challenging to visualize what features it is learning on. This effectively renders the neural net a “black box”, which poses a challenge when attempting to find the root cause of a misclassification, and gives an advantage to potential adversarial attacks. Furthermore, CNNs are highly influenced by dense pixel values which are not robust features compared to shapes of objects \cite{geirhos2018imagenettrained}. Thus, shapes of objects should be learned to allow generalizability and robustness of the model which goes hand-in-hand with transparency. A lack of model transparency and robustness will hinder its translation into a clinical setting.
\smallskip
\newline
\indent While CNNs have shown promise in ventricular segmentation of CMR images, a lack of transparency as to what is focused on during the segmentation will limit the translatability of such technology into a clinical setting. There is a dire need to improve the transparency of neural networks and we suggest one way to achieve transparency and robustness is to enforce the model to learn shape considerations. By affording a higher accuracy in segmentation and verifying that an algorithm is not perpetuating biases, a valuable tool can be created to help solve the challenges numerous clinicians face in medical image analysis.
\smallskip

\indent To the best of our knowledge, past attempts in incorporating shape information in medical imaging segmentation involve forming a new loss function \cite{10.1007/978-3-319-74113-0_2, 10.1007/978-3-030-00928-1_49, Chen_2019_CVPR}. Furthermore, limited works on model interpretability for medical imaging have been published. As such, the contributions of this work are:
\begin{itemize}
    \item The addition of a secondary stream that processes shape features of the image in parallel with the U-Net. Rather than constructing a new loss function, we suggest learning shape features can be built-in a model. Further, the output of the shape stream is a shape attention map that can be used for interpretability.
    \item The usage of spatial and channel-wise attention paths in the decoder block for interpretability of features the model learns at every resolution of the U-Net.
\end{itemize}
We evaluate our model on large public MRI ventricular volume estimation and segmentation datasets SUN09 \cite{sun09} and AC17 \cite{RN225} and demonstrate our method yields state-of-the-art results. We then provide an analysis of the spatial and shape attention maps to interpret the features our model learned. Consequently, our method not only yields strong performance, it is also interpretable at multiple resolutions.

\section{Related Works}
\indent In this section, we briefly review CNN-based segmentation architectures and methods related to this work.

\subsection{U-Net}
\indent U-Net consists of a contracting path that captures feature information as well as a symmetric expansive path that enables localization. Moreover, a U-Net uses skip connections from encoders to decoders of similar resolutions to pass high-resolution information throughout the network.
\smallskip
\newline
\indent In the original proposal of U-Nets, each block in the encoder is comprised of two successive normalized 3x3 convolutional layers. Then, a max-pooling layer with a 2x2 kernel of stride 2 is used to down-sample the image in order to obtain greater contextual and spatial information. In the up-sampling path, the same blocks in the encoder are used, except feature maps of similar resolutions from the down-sampling path are concatenated with the feature maps from the up-sampling path after being cropped and aligned to match dimensions before being filtered by a normalized 3x3 convolutional layer. Unlike the work of \cite{7298965} which uses element-wise summation of skip connections with up-sampling feature maps, \cite{10.1007/978-3-319-24574-4_28} simply concatenates them before applying 3x3 convolutions in the decoder.
In both of these works, the skip connections are used to obtain more fine-grained information that may have been lost during the intermediate stages. Following each decoder block, an up-sampling is applied to the feature maps to increase the resolution by 2. The original proposal uses interpolation but transpose convolutions can be used to learn the up-sampling parameters and have shown better results.

\subsection{DenseNet}
\indent DenseNets were originally proposed by Huang et al. \cite{Huang2016DenselyCC} and lead to significant improvements in state-of-the-art scores over previous models like ResNet \cite{RN227} and ResNeXt \cite{xie2016aggregated} in image classification tasks such as ImageNet. Furthermore, DenseNets achieved state-of-the-art scores while also having fewer parameters than previous models with similar performance (i.e. DenseNet-169 had better performance than ResNet-50 while also having approximately 50\% fewer parameters) due to feature reuse from the additional skip connections. A DenseNet is comprised of two main building blocks - a dense block followed by a transition block. A dense block contains many normalized 3x3 convolution layers where the outputs of each layer are concatenated with each of the feature maps entering the following layers to promote feature reuse. With $n$ layers in a dense block, there are $n!$ skip connections in the block. Each layer outputs a feature map with constant depth of $k$, so $n \times k$ channels exit the dense block. The transition block is composed of a normalized 1x1 convolution to reduce the depth of the feature maps followed by a $2 \times 2$ average pool with stride 2 to shrink the resolution by half.
\subsection{Attention in CNNs}
\indent Interpretability in computer vision comes in two flavours - post hoc analysis \cite{selvaraju2016gradcam, smilkov2017smoothgrad} and trainable attention. Post hoc analysis is employed after the model has been trained and typically uses gradient-based mechanisms to access the network's reasoning. However, these methods generally do not provide a way of interpreting the model's decision making at different stages of forward propagation and have been shown to be an unreliable source of interpretability \cite{kindermans2017unreliability}. Trainable attention are modules in the model that have learnable parameters. The purpose of these modules is to explicitly learn to reassign importance over some dimension. Jetley et al. \cite{RN222} proposed an attention estimator module that takes in intermediate feature maps at multiple stages of the network. Then, successive convolutions are applied to reduce the depth of these combined feature maps to 1 followed by a sigmoid or softmax to rescale each pixel value of the attention map to $[0,1]$. The attention map is then multiplied through all channels in the combined feature maps to re-calibrate the relative activations of each map. The authors showed empirically that popular models such as VGG-16 with their proposed attention mechanisms outperform the same model without it on cross-domain image classification datasets after trained on CIFAR-10 and CIFAR-100. Furthermore, the attention maps can be extracted to produce a visual heat-map that can be used for interpretability. The authors also showed these attention maps can be binarized and then used as maskings for weakly supervised semantic segmentation on image classification datasets. Schempler et al. \cite{schlemper2018attention} extended this attention estimator idea further by applying a similar module prior to each decoder block in their U-Net-based architecture tested on 3D CT image datasets.
\smallskip
\newline
\indent Squeeze and excitation blocks proposed by Hu et al. \cite{hu2017squeezeandexcitation} can also be thought of as trainable attention mechanisms. These modules are learned to explicitly redistribute the relative importance of each channel in a feature map according to each channel's global information captured by a global average pooling layer. Other authors have proposed U-Net variants that incorporate squeeze and excitation modules for medical image segmentation tasks and have achieved promising results. Guo et al. \cite{Guo2019GIANAPS} proposed adding a squeeze and excitation block prior to each decoder block such that it takes in the skip connection and the feature map from the previous lower resolution decoder block. Unfortunately, squeeze and excitation modules do not offer an obvious method of interpretability.
\smallskip
\newline
\indent Our work combines the recent advances of trainable channel-wise attention using squeeze and excitation modules proposed by Hu et al. \cite{hu2017squeezeandexcitation} and spatial attention using the attention estimator proposed by Jetley et al. \cite{RN222}. We show that incorporating attention along all dimensions yields strong results that are also inherently interpretable.

\smallskip
\subsection{Active Contour Loss for Learning Shapes} The active contour loss was recently introduced by Chen et al. \cite{Chen_2019_CVPR} in an attempt to incorporate shape considerations in the learning process. It was inspired by the work of \cite{journals/ijcv/KassWT88} on Active Contour Models (ACMs) and the authors proposed to incorporate area and size information into the loss function for more precise segmentation. The AC Loss is defined as,
            \begin{equation}
               Loss_{AC} = Length + \lambda Region
            \end{equation}
where \begin{math}Length\end{math} is the $L_2$ norm of the segmentation's contour and \begin{math}Region\end{math} compares the means of areas inside and outside the object using a pixel-wise $L_1$ norm loss. \begin{math}\lambda\end{math} is a fixed parameter that controls the balance between the above terms. \cite{Chen_2019_CVPR} showed that their model performed worse when $\lambda \to 0$ as only \begin{math}Length\end{math} contributed to the loss. Otherwise, their model's performance was invariant to the choice of \begin{math}\lambda\end{math}.

\indent Chen et al. combined this geometrically-constrained loss function with their U-Net with dense blocks and it outperformed other mainstream loss functions such as cross entropy loss. Although this loss function takes into account the geometric information of the areas being segmented, it only considers the length and area of inside and outside regions and does not actually consider the shape of the object being segmented. Furthermore, the formulation of the $Region$ term is a $L_1$ loss which has weak gradients compared to cross entropy. Using this loss function, we experimentally observed the phenomenon of slow learning and poor convergence.


\begin{figure*}[t]
\centering
\includegraphics[width=1.0\textwidth]{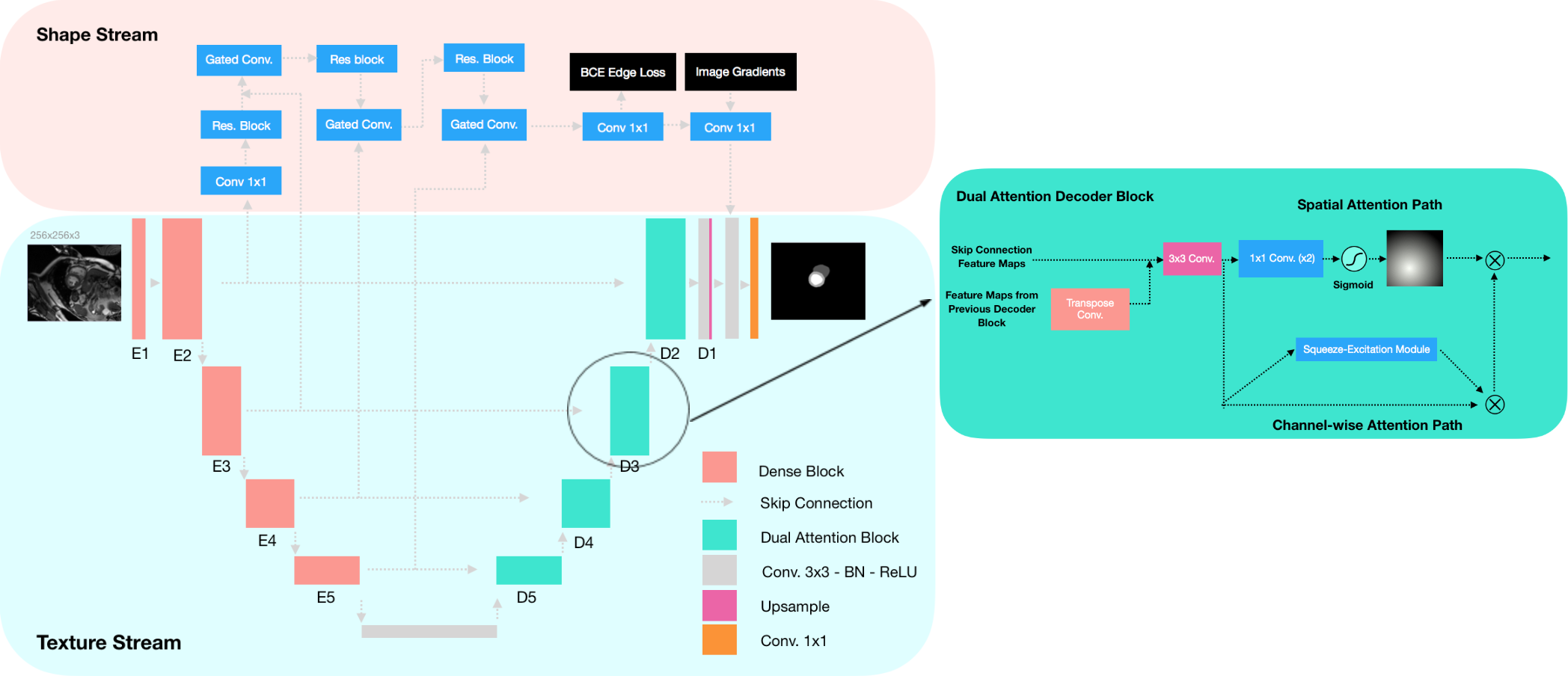}
\caption{Our proposed Shape Attentive U-Net. The proposed model is composed of two main streams - the shape stream that processes boundary information and the texture stream. The shape stream is composed of gated convolutional layers and residual layers. The gated convolutional layers are used to fuse texture and shape information while the residual layers are used to fine-tune the shape features.}
\end{figure*}

\section{Method}
We propose a new interpretable image segmentation architecture called Shape Attentive U-Net (SAUNet). A SAUNet is composed of two streams - the texture stream and the gated shape stream. The texture stream is the same structure as an U-Net, however the encoder is replaced with dense blocks from DenseNet-121 similar to the Tiramisu Network proposed by J\'egou et al. \cite{jgou2016layers} and the decoder block used is our proposed dual attention decoder block. Dense pixel information and features are learned through the texture stream but not shape features. Takikawa et al. \cite{takikawa2019gatedscnn} proposed Gated-SCNN, in which the authors first introduced the idea of a gated shape stream to help remove noise and produce finer segmentation on the Cityscape dataset. We propose that on top of producing finer segmentations, the gated shape stream gets the model to learn object shapes (see Figure 3 and \textit{supplementary materials}) and this is consequently interpretable. Hence, we propose the usage of a secondary stream on-top of our U-Net variant that processes shape and boundary information, the gated shape stream, explained in detail in Section 3.1.

\subsection{Gated Shape Stream}
The gated shape stream processes and refines relevant boundary and shape information using features processed by the encoder from the texture stream. The fusion of shape information from the shape stream flow with texture information is done by the gated convolutional layer.

\subsubsection{Gated Convolutional Layer} Let $C_{1\times 1}(x)$ denote the normalized 1x1 convolution function applied on feature map $x$, and let $R(x)$ denote the residual block function applied on feature map $x$. The residual block used is composed of two normalized 3x3 convolutions with a skip connection. The function $C_{1\times 1}(x)$ returns a feature map of the same spatial dimensions as $x$ but shrinks the number of channels down to one. The gated convolutional layer computes an attention map, $\alpha_l$, of boundaries by using information from the shape stream flow and the texture stream. Formally, denote the shape stream feature maps as $S_l$ and the texture stream feature maps as $T_t$ where $l$ denotes the layer number in our shape stream and $t$ indexes the encoder block the texture stream feature maps are outputted from. Bilinear interpolation is applied to $T_{t}$ if needed to match the dimensions of $S_{l}$. Since we want precise boundaries of the shape, no pooling layers should be used in the shape stream. We define each residual block as a layer for the shape stream. Then, $\alpha_l$ is computed as,

    \begin{equation}
        \alpha_{l} = \sigma(C_{1x1}(S_{l} || C_{1x1}(T_{t})))
    \end{equation}

\bigskip
\noindent where $\sigma$ is the sigmoid function and $||$ denotes the channel-wise concatenation of feature maps. $\alpha_{l}$ is stacked channel-wise to match the dimensions of $S_{l}$. Then, the output of the gated convolutional layer, $\hat{S_{l}}$, is the inputted shape stream feature map $S_{l}$ element-wise multiplied with $\alpha_{l}$,

    \begin{equation}
        \hat{S_{l}} = S_{l} {\scriptstyle\bigotimes} \alpha_{l}
    \end{equation}
\newline
\noindent where {\small$\bigotimes$} is the Hadamard product. The feature map of the next layer of the shape stream is computed as,

    \begin{equation}
        S_{l+1} = R(\hat{S_{l}})
    \end{equation}
\newline
\noindent and the same procedure to refine $S_{l+1}$ is applied (equations 2-3 but now $T_{t+1}$ is used).

\subsubsection{Output of Gated Shape Stream} The predicted class boundaries from the shape stream are deeply supervised to produce $L_{edge}$. $L_{edge}$ is the binary cross entropy loss between the ground truth class boundaries and the predicted class boundaries by the shape stream. Now, an objective of the model is to learn the shapes of the classes correctly. Since the entire gated shape stream is differentiable, the gradients propagate back to even the texture stream encoders. Intuitively, the texture stream encoders will learn some relevant shape information as well. The output of the gated shape stream is the predicted shape feature maps of the classes of interest concatenated channel-wise with the Canny edges from the original image. The output is then concatenated with the texture stream feature maps before the last normalized 3x3 convolution layer of the texture stream.

\subsection{Dual Attention Decoder Block}
The decoder module fuses feature maps outputted by the encoder from the skip connection along with the feature maps of lower resolution decoder blocks that capture more contextual and spatial information. Naturally, we would like to understand what features the model is detecting in these blocks to make the model less black-box. We propose the dual attention decoder block that is comprised of two new components after the standard normalized 3x3 convolution on the concatenated feature maps. The two new components are the spatial attention path for interpretability and a channel-wise attention path for improved performance as demonstrated by Hu et al. \cite{hu2017squeezeandexcitation}.

\subsubsection{Spatial Attention Path} Let $C$ denote the number of channels coming into the spatial attention path. Then, the spatial attention path is composed of a normalized 1x1 convolution followed by another 1x1 convolution. The first convolution reduces the number of channels to $\frac{C}{2}$ and the second convolution reduces the number of channels to 1. A sigmoid is applied to map the pixel values in the single channel into the range of $[0,1]$ to obtain $F'_{s}$. $F'_{s}$ is then stacked channel-wise $C$ times to obtain $F_{s}$. This is done to match the dimensions of the spatial attention path output, $F_{s}$, with the output from the channel-wise attention path, $F_{c}$, in order to perform element-wise multiplication.
\subsubsection{Channel Attention Path} The channel-wise attention path is comprised of a squeeze and excitation module that produces a scale coefficient in $[0, 1]$ for each channel from the skip connection. Each channel from the skip connection feature map is then scaled by their respective coefficient to obtain $F_{c}$.

\subsubsection{Channel and Spatial Attention} The output, $F$, of our proposed dual-attention decoder block is a fusion of channel and spatial attentions,
    \begin{equation}
        F = (F_{s} + 1) {\scriptstyle\bigotimes} F_{c}
    \end{equation}

Operator {\small$\bigotimes$} denotes the Hadamard product. The $ + 1$ is included so that the spatial attention originally in the range of $[0,1]$ can only amplify features rather than zeroing out features that may be valuable in later convolutions.

\subsection{Dual-Task Loss Function for Learning Shapes}
Our proposed objective function optimizes for precise segmentations and promotes the learning of shapes. We define each term of our objective in the following subsections.
\subsubsection{Cross Entropy Loss} Cross entropy loss is a commonly used loss function for image segmentation tasks. It is derived from the concept of entropy in information theory. Formally, we define the cross entropy between two distributions $p$ and $q$ to be the average number of bits required to map an event using an estimated probability distribution $q$ to the ground-truth distribution of $p$.
\smallskip
\newline
\indent For image segmentation, the cross entropy loss is calculated as the average cross entropy over all pixels. Let $\Omega$ denote the domain of all pixels with height $m$, width $n$, and $K$ classes. Let $y \in M_{m \times n \times K}(\{0,1\}) $ be the ground-truth one-hot matrix encoding the ground truth class of each pixel. Further, let $\hat{y} \in M_{m \times n \times K}([0,1])$ be a matrix of the predicted probabilities of each individual pixel. The cross entropy loss is defined as,

            \begin{equation}
                L_{CE}(\hat{y}, y) =
                \frac{1}{|\Omega|}\sum_{i}^{\Omega}-y_{i} log({\hat{y}_i}) - (1-y_i) log(1-\hat{y}_{i})
            \end{equation}

\smallskip
\subsubsection{Dice Loss} Dice loss is another common loss function used for image segmentation tasks as it measures the overlap and similarity between two sets. Given two countable sets A and B, the Dice coefficient is formally defined as,
                \begin{equation}
                    Dice(A, B) = \frac{2| A \cap B |}{| A \cap B | + | A \cup B |}
                \end{equation}
                \\
                \smallskip
Evidently, Dice(A, B) is maximized at 1 when $A = B$ and minimized at 0 when $A \cap B = \varnothing$ . The Dice loss function should be inverted and differentiable, so we define the Dice loss as the following,

                \begin{equation}
                    L_{Dice}(\hat{y}, y) = 1 - \frac{2}{K} \sum_{k=0}^{K-1} \frac{\sum_{i}^{\Omega}y^k \hat{y}_i^k}{\sum_{i}^{\Omega}y_{i}^k + \hat{y}_i^k}
                \end{equation}
                \\
                \smallskip
where $K$ is the total number of classes and $y_{i}^k$ denotes the $i^{th}$ pixel of the $k^{th}$ indexed class of matrix $y \in M_{m \times n \times K}(\mathbb{R})$.
\subsubsection{Dual Task Loss}
\indent We propose the loss function used to be composed of the segmentation loss and the shape stream boundary loss. Let $L_{CE}$ and $L_{Dice}$ denote the cross entropy loss and Dice loss of the predicted segmentation respectively. Let $L_{Edge}$ denote the binary cross entropy loss of the predicted shape boundaries. Then, our total loss, $L_{total}$, is defined as,

        \begin{equation}
            L_{total} = \lambda_{1}L_{CE} + \lambda_{2}L_{Dice} + \lambda_{3}L_{Edge}
        \end{equation}
\newline
where $\lambda_{1}$, $\lambda_{2}$, and $\lambda_{3}$ are hyper-parameters to weigh each measure. In our experiments, we find setting $\lambda_{1}$ = $\lambda_{2}$ = $\lambda_{3}$ = 1 works well. Moreover, we found that using $L_{CE}$ without $L_{Dice}$ and vice versa results in the model not learning well and quickly. Hence, both $L_{CE}$ and $L_{Dice}$ are used. Intuitively, the model learns to predict individual pixel values correctly through $L_{CE}$ and also learns to consider overlap through $L_{Dice}$.


\section{Experiments}
\smallskip
For this study, we conduct our experiments on the SUN09 and AC17 segmentation datasets. We present our results in the following sections. The experiments were completed on one NVIDIA Quadro RTX 5000 GPU with 16GB of memory. For each experiment, each image slice was z-score normalized. The following data augmentations were performed during runtime: rotations in $[-\pi, \pi]$, horizontal and vertical flips with 50\% chance, elastic deformations, and gamma shifts with a factor sampled from $[0.5, 2.0]$ uniformly distributed. All the code used is publicly available on our GitHub: \textit{https://github.com/sunjesse/shape-attentive-unet}

\subsection{SUN09 Left Ventricle Segmentation Dataset}
The SUN09 dataset contains separate training datasets for each of the two classes - the endocardium and the epicardium. The dataset for endocardium segmentation consists of 260 2D MRI slices. Likewise, the dataset for epicardium segmentation consists of 135 2D MRI slices. Each slice was center cropped to a resolution of 128px by 128px. Different datasets were given for each of the two classes so two models were trained for SUN09. RAdam \cite{liu2019variance} optimizer was used with $\beta_1$ = 0.9, $\beta_2$ = 0.999, along with weight decay value of 1E-4, and initial learning rate of 7E-4 exponentially decayed with parameter 0.99. Each model was trained for 120 epochs with a batch size of 4.

\subsection{AC17 Bi-Ventricular Segmentation Dataset}
\indent The AC17 dataset contains 200 volumes of varying resolution MRI scans from 100 unique patients from the University Hospital of Dijon \cite{RN225}. Each volume has a depth of 8-20 slices. To account for varying scales of each image, the pixels of every slice and their respective mask were resampled to 1.25mm by 1.25mm without changing the aspect ratio. The original image used bilinear interpolation while the segmentation mask used nearest-neighbour interpolation to avoid continuous values. The volumes were not rescaled along the z-axis. A center-crop of resolution 256px by 256px was made, and zero-padding was applied if necessary. We noticed that the minimum value of some slices was greater than zero; for consistency with the zero-padding made, the minimum value of every slice was subtracted before being zero-padded so that the minimum pixel intensity of every slice is 0 prior to adding the zero-pad and normalizing. The training dataset was randomly shuffled and split into an 80:20 train-validation split in which we seeded. The model was trained using RAdam for 180 epochs with $\beta_{1}$ = 0.9, $\beta_{2}$ = 0.999, along with weight decay value of 1E-4 and initial learning rate of 5E-4 exponentially decayed with parameter 0.99. Two-dimensional batch normalization was used with a batch size of 10. We report and submit the results to the test set of the convergence epoch.


\section{Experimental Results}
\indent We present our results on the SUN09 left ventricle segmentation dataset and the AC17 left/right ventricle and myocardium segmentation dataset. The Dice Coefficient metric is used for evaluation consistent with other benchmarks and works.
\subsection{SUN09}
 For SUN09, training our proposed model for 120 epochs took 1.5 hours for each class' dataset. No hyper-parameter tuning was done for our model trained with the shape stream. Our scores along with the previous top five scores are reported in Table 1.
\begin{table}[H]
\centering
\begin{tabular}{lcc}
\multicolumn{3}{c}{} \\
\cline{1-3}
Model   & Endocardium & Epicardium  \\
\hline
Tran \cite{tran2016fully} & 0.92 & 0.96 \\
Curiale et al. \cite{curiale2017automatic}  & 0.90 & 0.90 \\
Yang et al. \cite{8120080} & 0.93 & 0.93 \\
Romaguera et al. \cite{10.1117/12.2253901} & 0.90 & - \\
Avendi et al. \cite{avendi2015combined} & 0.94 & - \\
\hline
\textbf{SAUNet (Ours)}  & \textbf{0.952}      & \textbf{0.962}       \\
\hline
\end{tabular}
\caption{Test set Dice scores for SUN09. Only the previous top five works with the highest reported accuracies (Dice scores) are listed. No hyper-parameter tuning was required for our SAUNet model.}
\end{table}

\subsection{AC17 Segmentation}
For AC17, training our proposed model for 180 epochs took 3.5 hours. The Dice scores for the left ventricle, right ventricle, and myocardium segmentation structures are presented in Table 2 along with the scores from previous published works on AC17. The results presented are for the model trained with RAdam and a learning rate of 5E-4.

\begin{table}[H]
\centering
\setlength{\tabcolsep}{10.5pt}
\begin{tabular}{lccc}\toprule
& \multicolumn{1}{c}{LV} & \multicolumn{1}{c}{RV} & \multicolumn{1}{c}{MYO} \\
\cline{1-4}
\hline
Wolterink et al. \cite{wolterink} & 0.930 & 0.880 & 0.870 \\
Patravali et al. \cite{jay} & 0.925 & 0.845 & 0.870  \\
Khened et al. \cite{khened}  & 0.920 & 0.870 & 0.860 \\
Ilias et al. \cite{inbook} & 0.905& 0.760 & 0.785  \\
Jang et al. \cite{jang} & 0.938  & 0.890 & 0.879 \\
\hline
\textbf{SAUNet (Ours)} & \textbf{0.938} & \textbf{0.914} & \textbf{0.887}  \\
\hline
\end{tabular}
\caption{AC17 test set results. Our proposed method yields state-of-the-art results for end-to-end models in all classes in terms of Dice score. Notably, our performance in the right ventricle class is significantly better than previous works, and this is perhaps due to the help of the shape stream learning the irregular shape of the right ventricle.}
\end{table}

\subsection{Ablation Studies}
\indent To test the effectiveness of our proposed gated shape stream, we conducted ablation studies on our model using the AC17 dataset. Results reported in Table 3 are from our validation set of our proposed model with and without the shape stream. Both models were trained with learning rate 5E-4 and RAdam. The results reported are the models taken at the convergence epoch. Both models were then evaluated on the test set. Figure 2 shows some cases from the test set where the model with the shape stream visibly outperforms its counterpart.

\begin{table}[H]
\begin{tabular}{lccc}
\multicolumn{3}{l}{} \\
\cline{1-4}
Model   & RV & MYO & LV\\
\hline
SAUNet w/o Shape Stream & 91.40 &  88.31 & 94.96\\
SAUNet w/ Shape Stream  & \textbf{93.11} &   \textbf{88.64}  & \textbf{95.71}\\
\hline
\end{tabular}
\caption{The Dice scores in percentage of our model with and without the proposed shape stream evaluated on our AC17 validation set split. The accuracy of the right ventricle class increased a significant amount as the irregular shape of the right ventricle was learned using the shape stream.}
\end{table}

\begin{figure}[H]
\includegraphics[width=0.48\textwidth]{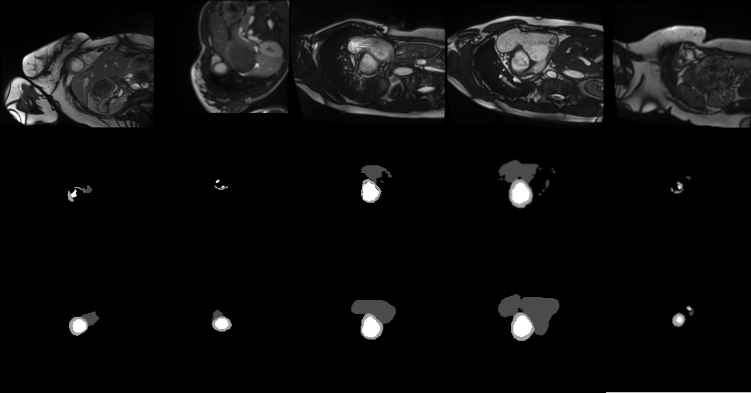}
\caption{Top: original MRI image. Middle: model without shape stream prediction. Bottom: model with shape stream prediction. Higher quality segmentations with the correct shapes of the classes are made using the shape stream. }
\end{figure}

\subsection{Robustness of Shape Stream}
Intuitively, shapes are more robust features than textures. To evaluate the effectiveness of our dual-task objective of concurrently learning shapes through our proposed shape stream, we set up the following experiment: train the model with and without the shape stream on only SUN09 data, and then evaluate each model on our AC17 validation set split. The SUN09 dataset contains only labelled segmentations for the left ventricle, so only the left ventricle class was considered when evaluating on AC17. We trained both models with RAdam, initial learning rate of 5E-4, and a batch size of 4 for 120 epochs. The mIoU scores are reported in Table 4.

\begin{table}[H]
\begin{tabular}{lccc}
\multicolumn{3}{l}{} \\
\cline{1-4}
Model   & SUN09 Train & AC17 Val. & Drop\\
\hline
w/o Shape Stream & 90.20 &  79.33 & -10.87\\
w/ Shape Stream  & \textbf{90.84} &   \textbf{81.73}  & \textbf{-9.11}\\
\hline
\end{tabular}
\caption{The mIoU scores in percentage of training on SUN09 and testing on our AC17 validation set split. The percentage drop of using the model with the shape stream is 1.76\% less than otherwise suggesting shapes are robust features that generalize well and hence should be learned. The scores under SUN09 Train is the SUN09 test set mIoU score.}
\end{table}
\bigskip

\begin{figure*}[t]
  \centering
  \includegraphics[width=1.0\textwidth]{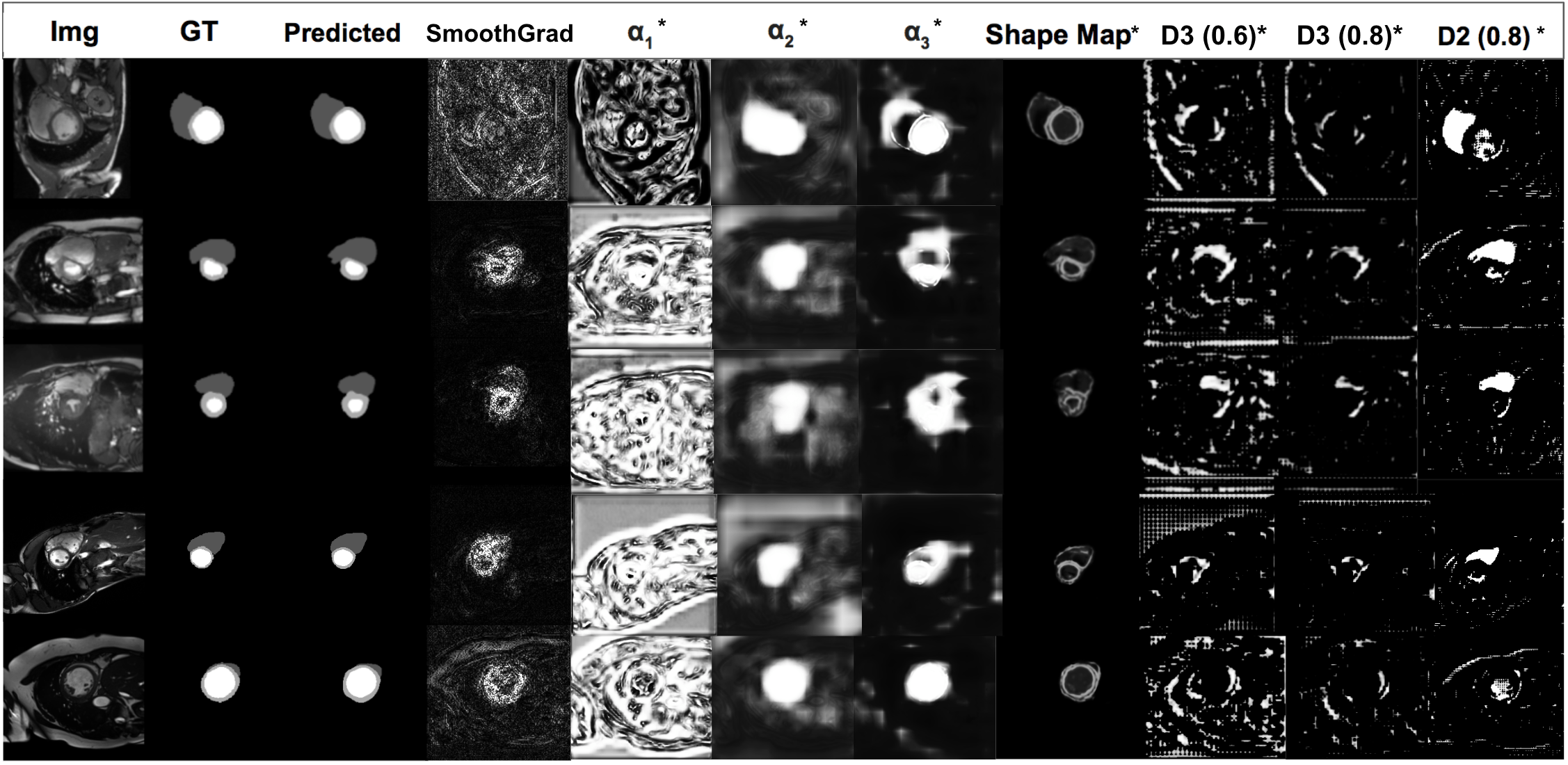}
  \caption{Our models attentions. * denotes proposed method. $\alpha_{l}$ is $l^{th}$ shape attention map, $DX(Y)$ is $X^{th}$ decoder block threshold of $Y$.}
\end{figure*}

\section{Interpretability}
\indent The learned shape and spatial attention maps are extractable from our model. The spatial attention maps can be used to interpret the regions of high activation for each decoder block while the learned shape maps can be used to deduce that the model has learned the correct shape of the classes of interest. Figure 3 contains the spatial attention map of the last two dual-attention decoder blocks, intermediate shape stream attentions $\alpha_{l}$, the final shape attention map, and saliency maps generated using SmoothGrad \cite{smilkov2017smoothgrad}.

\indent Higher-resolution decoder blocks learned the feature of the right ventricle primarily according to column 11 (\textit{D2(0.8)}) where the maps are thresholded by 0.8. In the last row, the right ventricle in the image is not visible so high attention was placed on the left ventricle. The right ventricle seems to have the highest attention and priority as its shape is unique and hence serves as a robust feature for localization. The attention maps from decoder block 3 are at a lower resolution level and according to column 10 (\textit{D3(0.8)}) where the spatial attention maps are thresholded by 0.8, fewer areas of high activation are present compared to decoder block 2. Lower resolution blocks process more global information, so similar attention is placed among many regions of the image. To support this claim, column 9 (\textit{D3(0.6)}) presents the same attention maps but thresholded by 0.6. Not only traces of attention around the right ventricle are apparent, many other regions like the brim of the structure are also in focus. Furthermore, the shape stream narrows down on the shape of the structures of interest evident in columns 5-8.
\smallskip

Our method of interpretability has a few advantages over post hoc analysis methods like saliency maps \cite{selvaraju2016gradcam, smilkov2017smoothgrad}. SmoothGrad \cite{smilkov2017smoothgrad} is highly regarded currently as it overcomes the pitfalls of previous gradient-based saliency methods through averaging gradients. However, each image's saliency map requires 25-50 forward-backward passes to generate. Using SmoothGrad, it took 24 minutes to generate the saliency maps of all 384 images in our validation set, while it only took 20 seconds using our proposed method. Furthermore, our proposed method offers saliency methods at different levels and resolutions in which SmoothGrad and many other gradient-based methods do not provide. Conversely, GradCam \cite{selvaraju2016gradcam} offers a solution to view the activation of every convolutional layer. However, GradCam requires one forward-backward pass post hoc per saliency map while our proposed method generates all of the multi-resolution saliency maps on the initial forward pass.

\section{Conclusion}
\indent In this work, we present a new inherently interpretable medical image segmentation model called Shape Attentive U-Net. Our proposed method is able to learn robust shape features of objects via the gated shape stream while also being more interpretable than previous works via built-in saliency maps using attention. Furthermore, our method yields state-of-the-art results on the large public bi-ventrciular segmentation datasets of SUN09 and AC17. Through this work, we hope to take a step in making deep learning methods clinically adoptable and hopefully inspire more work on interpretability to be done in the future.
\newpage
{\small
\bibliographystyle{ieee_fullname}
\bibliography{egbib}
}

\end{document}